\documentclass[aps,pra,reprint,10pt,superscriptaddress,showpacs,amssymb,amsfonts,floatfix,nobalancelastpage]{revtex4-1}
\usepackage{amsmath,amssymb,latexsym,epsfig,bm,subfigure,booktabs}

\usepackage[usenames,dvipsnames,table]{xcolor}\usepackage[colorlinks,bookmarks=false,citecolor=RoyalBlue,linkcolor=BrickRed,urlcolor=RoyalBlue]{hyperref}
\usepackage{verbatim}

\def\beq{\begin{equation}}
\def\eeq{\end{equation}}

\def\ket#1{\vert #1 \rangle}

\newcommand{\bfm}{\mathbf}
\newcommand{\up}{\uparrow}
\newcommand{\down}{\downarrow}

\def\bq{\bfm{q}}
\def\br{\bfm{r}}

\begin{document}

\title{Spectroscopic signatures of crystal momentum fractionalization}

\author{Andrew M.~Essin}
\affiliation{Department of Physics, 390 UCB, University of Colorado,
Boulder CO 80309, USA}
\affiliation{Institute for Quantum Information and Matter, California Institute of Technology, Pasadena CA 91125, USA}
\author{Michael Hermele}
\affiliation{Department of Physics, 390 UCB, University of Colorado,
Boulder CO 80309, USA}
\date{\today}

\begin{abstract}
We consider gapped ${\mathbb Z}_2$ spin liquids, where spinon quasiparticles may carry fractional quantum numbers of space group symmetry.  In particular, spinons can carry fractional crystal momentum.  We show that such quantum number fractionalization has dramatic, spectroscopically accessible consequences, namely enhanced periodicity of the two-spinon density of states in the Brillouin zone, which can be detected via inelastic neutron scattering.  This effect is a sharp signature of certain topologically ordered spin liquids and other symmetry enriched topological phases.  Considering square lattice space group and time reversal symmetry, we show that exactly four distinct types of spectral periodicity are possible.
\end{abstract}


\maketitle

Quasiparticle excitations with fractional quantum numbers are a striking feature of certain quantum phases of matter.  Shot-noise and local tunneling experiments have observed fractional electric charge $Q_e = e/3$ in fractional quantum Hall systems~\cite{depicciotto97,saminadayar97,Martin2004}, where $e$ is the electron charge.  Inelastic neutron scattering experiments have observed $S=1/2$ excitations in spin chains~\cite{Tennant1993,Zaliznyak2004}; such excitations are fractional, because any local process (e.g., spin flips relative to the ground state) can only create \emph{integer}-spin excitations.  These results point out that, while in many cases exotic quantum phases are characterized by subtle properties that are challenging to probe, fractional quantum numbers can offer sharp, experimentally accessible signatures.

While charge and spin fractionalization are familiar possibilities, it is 
becoming clear that quantum numbers of \emph{space group} symmetry can also fractionalize~\cite{wen02, Wen02.2, Essin2013}, although physical consequences largely have yet to be illuminated.  We show that 
space-group fractionalization has dramatic, surprising signatures accessible to spectroscopic probes.  In particular, after discussing what it means for crystal momentum to be fractional, we explain how this leads to an increased periodicity of the excitation spectrum in momentum.  
We also describe similar signatures for other types of space group symmetry fractionalization.  These phenomena have the potential to play an important role in the experimental identification of new exotic phases of matter.

We focus on two-dimensional gapped $\mathbb{Z}_2$ spin liquids \cite{chakraborty89,read91,wen91,balents99,senthil00,moessner01a, moessner01b,balents02,kitaev03} with $S = 1/2$ spinon quasiparticles.  These are perhaps the simplest quantum spin liquids, and there is recent evidence these phases occur in  fairly realistic Heisenberg models~\cite{yan11,lwang11,jiang12,depenbrock12,jiang12b,mezzacapo12}.  The gapped spinons can have either bosonic or fermionic statistics, and also have $\theta = \pi$ mutual statistics with gapped bosonic visons ($\mathbb{Z}_2$ fluxes). Crucially, single spinons cannot be created by any local process, but spinon pairs can be created locally.   Recently, for gapped $\mathbb{Z}_2$ spin liquids and other related phases, we developed a  general description of quantum number fractionalization as a robust distinguishing characteristic of a quantum phase of matter~\cite{Essin2013}, which is the foundation for the results presented here.  

Results similar to those presented here will hold for symmetry-enriched topological phases \cite{wen02,wen03,fwang06,kou08,kou09,yao10,huh11,gchen12,levin12,cho12,Essin2013,Mesaros2013,Hung2013,Lu2013,Hung2013b,Xu2013,Hung2013c,Lu2013b,Huang2013} (of which gapped $\mathbb{Z}_2$ spin liquids are an example), characterized by an interplay between symmetry and topological order.  An interesting question for future study is whether our results extend---likely in a weaker form---to gapless spin liquids.

We will be interested in the two-spinon $S=1$ continuum.  This contributes to the dynamic spin structure factor $S(\bq,\omega)$ and can be probed with inelastic neutron scattering.  The continuum has density of states ${\cal N}(\bq,\omega)$ as a function of crystal momentum $\bq$ and energy $\hbar \omega$, with low-frequency edge $\Omega(\bq)$.  We show that crystal momentum fractionalization appears as an enhanced periodicity of ${\cal N}(\bq,\omega)$ in the Brillouin zone.  
Other types of fractional space group symmetry also have striking spectral periodicities.  
It should be noted that $S(\bfm{q},\omega)$ depends 
not only on ${\cal N}(\bq,\omega)$ but also 
on matrix elements not expected to exhibit the same periodicity, but $\Omega(\bq)$ is directly observable in $S(\bq,\omega)$.  While we focus on spinons for concreteness, identical considerations apply to the two-vison continuum, which could potentially be measured as well.

For simplicity, we assume throughout that single spinons are infinitely long-lived.  In practice, finite-lifetime effects will arise due to scattering off crystalline disorder and decay of single spinons into other excitations.  For our results, disorder
needs to be weak enough that space group symmetry is present to a good approximation.  We expect the principal effect of weak disorder to be some 
broadening of states in crystal momentum; future studies of such effects could be important for detailed comparisons to experimental data.  The decay of single spinons is constrained by their quantum numbers and the fusion rules of $\mathbb{Z}_2$ topological order, and in many circumstances there are infinitely long-lived single spinon excitations.  
For example, if spinons are lower in energy than visons and spinon-vison bound states, the spinons in some finite energy range above the spinon gap cannot decay.   
In general, spinon decay becomes an issue only when it significantly mixes the two-spinon continuum with other excitations.

We briefly comment on the relation between our results and previous work.  The standard tool used to study spatial symmetry in spin liquids is the projective symmetry group (PSG) approach \cite{wen02} for parton mean-field theories.  This approach describes the action of symmetry on mean-field spinons, and gives a symmetry-based classification of mean-field states, but it is not clear whether this structure has a robust meaning beyond mean-field theory.  This issue is important, partly because there is no reason to believe all spin liquids are perturbatively accessible starting from a parton mean-field theory.  In Ref.~\onlinecite{Essin2013}, we addressed this issue for gapped $\mathbb{Z}_2$ spin liquids, replacing the mean-field notion of PSG with the related notion of fractionalization class, which is a robust property of a quantum phase of matter.  Ref.~\onlinecite{wen02} calculated $\Omega(\bq)$  for several mean-field spin liquids with different PSGs.  Ref.~\onlinecite{Wen02.2} pointed out enhanced spectral periodicity---at the mean-field level---for ``Z2B'' spin liquids, corresponding to fractional crystal momentum or Type Q periodicity in our terminology. 
Ref.~\onlinecite{wenbook9.7} further mentioned what we will call Type D periodicity, although entirely in the context of gapless spin liquids and, again, at the mean-field level.
Compared to these prior works, our results identify new types of enhanced spectral periodicity, and make it clear that this periodicity is a robust property of a phase of matter.

\emph{Fractional crystal momentum}.  We begin by summarizing the necessary results from Ref.~\cite{Essin2013}, first considering the simple case of translation symmetry alone.  The translation symmetry of a two-dimensional crystal is generated by two elementary translations $T_x$ and $T_y$, which commute and thus satisfy
\beq
T_x T_y T^{-1}_x T^{-1}_y = 1 \text{.} \label{eqn:tcomm}
\eeq
This relation completely defines the translation group and holds in the entire many-body Hilbert space of the system of interest.  Energy eigenstates have definite crystal momenta $\bq$, namely $T_{\mu} | \bq \rangle = e^{i q_{\mu}} | \bq \rangle$, where $\mu = x,y$.  Crystal momentum is thus defined within the Brillouin zone $q_x, q_y \in [-\pi, \pi)$.

To consider the action of symmetry on spinon quasiparticles, we introduce translation operators $T^s_x$ and $T^s_y$ acting on a single spinon.  These operators locally translate the spinon against the translation-invariant background of the ground state.  (See Ref.~\cite{Essin2013} for more details on the definition and meaning of such operators.)  Because spinons come in pairs, we have
\beq
T_x^s T_y^s T_x^{s-1} T_y^{s-1} = \pm 1, \label{eqn:fcm}
\eeq  
where consistency with Eq.~(\ref{eqn:tcomm}) requires  the right-hand side to be the same for every spinon quasiparticle.  Thus translations may commute or anticommute when acting on single spinons, and \emph{anticommuting translations can be viewed as fractionalization of crystal momentum}.
Below, we clarify this terminology by defining a fractional crystal momentum quantum number.  For now, we draw an analogy with fractionalization of a $U(1)$ charge $\hat{Q}$, where $e^{2 \pi \hat{Q}} = 1$ on (integer-charged) many-body wavefunctions, but  $e^{2 \pi \hat{Q}_p} = e^{2 \pi i q_p}$ where $\hat{Q}_p$ is a charge operator for a single quasiparticle $p$ carrying fractional charge $q_p$.  Note that noncommuting translations are familiar in the magnetic translation group symmetry of charged particles in a uniform magnetic field, but that situation is physically distinct from the one considered here.

We are interested in the continuum of two-spinon states in the case of fractionalized crystal momentum, so we take the minus sign in Eq.~(\ref{eqn:fcm}).  Consider a scattering state of two spinons (and energy eigenstate) $\ket{a} = \ket{\bq_a ;z_a}$, with total crystal momentum  $\bq_a$, and where $z_a$ represents other labels needed to specify the state.  Using 1 and 2 to label the two spinons, translation symmetry acts on this state by
\begin{equation}
T_{\mu} \ket{a} = T^s_{\mu}(1) T^s_{\mu}(2) \ket{a}  \text{.}
\end{equation}
From $\ket{a}$, we can obtain three more states by applying single-spinon translation operators to spinon 2, namely
\beq
\ket{b} = T_x^s(2) \ket{a}, \quad \ket{c} = T_y^s(2) \ket{a}, \quad 
\ket{d} = T_x^s(2) T_y^s(2) \ket{a}.
\eeq
All these states are eigenstates with the same energy as $\ket{a}$; the two spinons do not interact in this scattering state (in the thermodynamic limit), so that shifting only one of the particles is a good symmetry.  Moreover, these states have distinct crystal momenta; using Eq.~(\ref{eqn:fcm}) it is straightforward to show 
\beq
\bq_b = \bq_a + (0,\pi), \quad \bq_c = \bq_a + (\pi,0), \quad \bq_d = \bq_a + (\pi,\pi) \text{.}
\eeq
That is, any two-spinon scattering state with crystal momentum $\bq_a$ is part of a degenerate multiplet of four states, with crystal momenta as given above, and therefore 
\begin{eqnarray}
{\cal N}[\bq,\omega] &=& {\cal N}[\bq + (\pi,0), \omega]   = {\cal N}[\bq + (0,\pi),\omega]  \label{eqn:typeq} \\ &=& {\cal N}[\bq + (\pi,\pi),\omega] \text{,} \nonumber
\end{eqnarray}
where the same enhanced periodicity holds for $\Omega(\bq)$.

Although simple to derive, this is a remarkable result.  A neutron-scattering experiment will see a spectrum that repeats four times within the Brillouin zone.  Such behavior would also occur with spontaneously broken translation symmetry (e.g., valence bond crystal order), but would be associated there with an elastic Bragg peak and a finite-temperature phase transition. Here, the periodicity is an intrinsic, generic property of a symmetric state. Without fractionalization of crystal momentum (or translation-breaking long-range order), such a periodicity would require extreme (and extremely unlikely) fine-tuning of parameters.

Another perspective on this result, which also clarifies the terminology ``fractionalization of crystal momentum,'' is provided by assigning crystal momentum to individual spinons.  In general, this assignment means that we label states by irreducible representations of the translation group.  When $\sigma_{txty} = 1$, all irreducible representations of the translation group are one-dimensional and have $T^s_\mu = e^{i k_\mu}$, so spinons carry ordinary crystal momentum $\bfm{k}$.  When $\sigma_{txty} = -1$, all irreducible representations, up to similarity transformation, are of the form
\begin{equation}
T_x^s = e^{i {\kappa}_x} \tau^x, \; T_y^s = e^{i \kappa_y} \tau^z, \label{eqn:2dirrep}
\end{equation}
where $\tau^x$ and $\tau^z$ are $2 \times 2$ Pauli matrices.  Single spinons are now labeled by a fractional crystal momentum $\boldsymbol{\kappa}$, where $\kappa_x$ and $\kappa_y$ are only defined modulo $\pi$, because the signs of $e^{i \kappa_x}$ and $e^{i \kappa_y}$ can be changed independently by suitable unitary transformations.  The enhanced periodicity of Eq.~(\ref{eqn:typeq}) can be seen directly from Eq.~(\ref{eqn:2dirrep}), by observing that the tensor product of two multiplets with fractional crystal momenta $\boldsymbol{\kappa}$, $\boldsymbol{\kappa}'$ gives operators
\beq
T^{2s}_x = e^{i(\kappa_x+\kappa_x')}\tau^x \otimes \tau^x, \quad
T^{2s}_y = e^{i(\kappa_y+\kappa_y')}\tau^z \otimes \tau^z,
\eeq
which commute and whose eigenstates have crystal momenta $\bq$, $\bq + (\pi,0)$, $\bq + (0,\pi)$, $\bq + (\pi,\pi)$, where $\bq = \boldsymbol{\kappa} + \boldsymbol{\kappa}'$.

\begin{figure}
\includegraphics[height=1in]{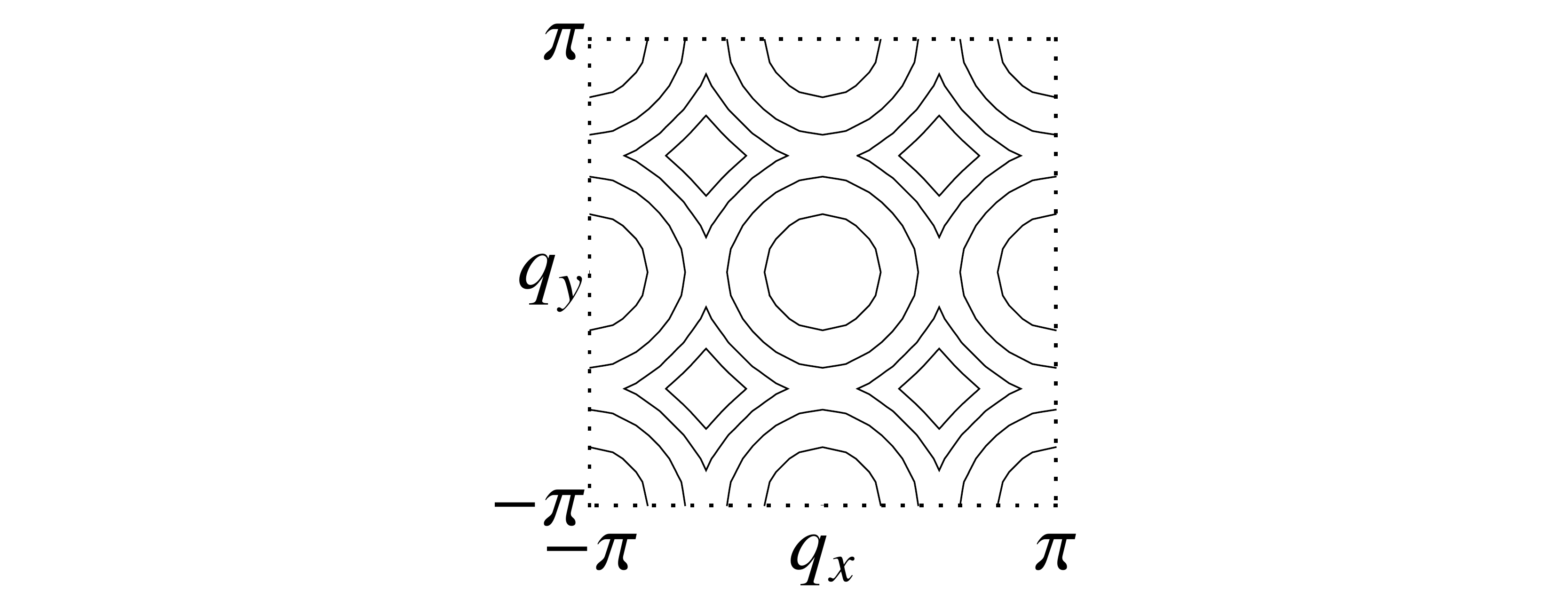}
\caption{ \label{fig:txtytor} 
Contour plot of $\Omega(\bfm{q})$ for the toric code model of Eq.~(\ref{eq:tc}), where spinons have fractionalized crystal momentum, in first-order perturbation theory.  Type Q spectral periodicity is present.
}
\end{figure}

As an example, we compute the excitation spectrum of a perturbed toric code Hamiltonian~\cite{kitaev03}.  The model has spin-1/2 moments on the links $\ell$ of a square lattice, with Hamiltonian
\beq \label{eq:tc}
H = -K_e \sum_s A_s - K_m \sum_p B_p - h \sum_\ell \sigma_\ell^z,
\eeq
where $s$ and $p$ label vertices and faces of the lattice and
$A_s = \prod_{ \{ \ell | s \in \partial \ell \} } \sigma^x_{ \ell}$, 
$B_p = \prod_{\ell \in p} \sigma^z_{\ell}$.  
We choose $|K_m| \gg |K_e|$ so that the lowest-energy excitations are pairs of $e$-particles (``spinons''), with each spinon lying on a vertex $s$ with $A_s = -1$.  
For $h \neq 0$ the spinons are able to move; when $K_m < 0$ this motion occurs in a background ``magnetic'' flux of $\pi$ per unit cell.  At first order degenerate perturbation theory in $h$, working in a convenient gauge with a two-site unit cell, the single-spinon dispersion is
\beq
E_{1,\pm}(\boldsymbol{\kappa}) = 2K_e \pm 2h \sqrt{\cos^2 \kappa_x + \cos^2 \kappa_y} \text{.}
\eeq
Single-spinon energies add to give the energies of two-spinon scattering states $E_2(\bq) = E_1(\bfm{q}-\bfm{k}) + E_1(\bfm{k})$, allowing us in particular to determine
$\Omega(\bq)$, shown in Fig.~\ref{fig:txtytor}.  Notice that the minimum at $(0,0)$ is repeated at $(0,\pi)$, $(\pi,0)$, and $(\pi,\pi)$. 

\emph{Fractionalization of space group symmetry.}  If translations are the only symmetries, the only possible symmetry fractionalization is that described above.
When other crystal symmetries are present, the number of  fractionalization classes grows \cite{Essin2013}, and we will now show that there is a corresponding increased richness of momentum-space signatures.  In particular, considering square lattice space group and time reversal symmetry, we find four distinct types of spectral periodicity, which we now name and describe:
\begin{itemize}
\item Type Q spectral periodicity is the  quadrupling of the spectrum in momentum space discussed above [Eq.~(\ref{eqn:typeq})].  
\item  Type D is a spectral doubling, ${\cal N}(\bq,\omega) = {\cal N}[\bq + (\pi,\pi),\omega]$.
\item  Type D$1d$ is exposed upon defining a reduced density of states depending only on $q_x$ (or $q_y$), ${\cal N}_x(q_x, \omega) = \int dq_y \, {\cal N}(\bq, \omega)$. This satisfies ${\cal N}_x(q_x,\omega) = {\cal N}_x(q_x + \pi, \omega)$.  In an experiment, D$1d$ spectral periodicity would be manifest in the two-spinon leading edge of $\int\! S(\bfm{q},\omega) dq_x$.
\item Type O refers to the ordinary restrictions of space group and time reversal on ${\cal N}(\bq,\omega)$, which are always present for a symmetric spectrum.
\end{itemize}
There is a simple hierarchical relation among these types of spectral periodicity; for instance, it is clear that type Q implies type D.  In general,
\begin{equation}
{\rm Q} \implies {\rm D} \implies {\rm D}1d \implies {\rm O} \text{.}
\end{equation}

The symmetries of the square lattice ($p4m$) are generated by translation $T_x$ and by two reflections:
\beq
P_x: \; (x,y) \rightarrow (-x,y), \quad P_{xy}: \; (x,y) \rightarrow (y,x),
\eeq
with $T_y = P_{xy} T_x P_{xy}$.
We also consider antiunitary time reversal $\mathcal{T}$.  The symmetry group can be specified by 10 relations involving these generators \cite{Essin2013,note1}; each relation can be augmented with a minus sign as in Eq.~(\ref{eqn:fcm}), leading to $2^{10}$ fractionalization classes for spinons \cite{Essin2013}.  We focus on the four relations involving $T_x$, as only these affect the spectrum as resolved by $\bq$:
\begin{alignat}{2} \label{eq:rels}
T_x^s T_y^s T_x^{s-1} T_y^{s-1} &= \sigma_{txty} 
& T_x^s P_x^s T_x^s P_x^{s-1} &= \sigma_{txpx} \notag\\
T_y^s P_x^s T_y^{s-1} P_x^{s-1} &= \sigma_{typx} 
&\quad
\mathcal{T}^s T_x^s \mathcal{T}^{s-1} T_x^{s-1} &= \sigma_{Ttx} ,
\end{alignat}
where each $\sigma_i = \pm 1$, giving sixteen combinations.  
Due to the six relations not considered, each choice of the $\sigma_i$ actually corresponds to $2^6$ fractionalization classes, all with the same spectral periodicity.  The results of the following discussion are summarized in Table~\ref{tab}, which gives the strongest periodicity present for each choice of $\sigma_i$.
\begin{table}
\caption{Type of spectral periodicity, as determined by the parameters of the 
fractionalization class for square lattice space group and time reversal symmetry. The rightmost column gives the \emph{strongest} type of spectral periodicity present. A dash (\textemdash) indicates that the value of the specified parameter $\sigma_i$ does not affect on the type of spectral periodicity.}
\label{tab}
\begin{ruledtabular}
\begin{tabular}{c c c c c}
$\sigma_{txty}$ & $\sigma_{tS}$ & $\sigma_{typx}$ & $\sigma_{txpx}$ & Spectral periodicity \\
\noalign{\vskip 1mm}
\hline
\noalign{\vskip 0.5mm}
-1 & \textemdash & \textemdash & \textemdash & Q \\
1 & -1 & \textemdash & \textemdash & D \\
1 & 1 & -1 & \textemdash & D$1d$ \\
1 & 1 & 1 & \textemdash & O \\
\end{tabular}
\end{ruledtabular}
\end{table}

For the eight cases with $\sigma_{txty} = -1$ (fractional crystal momentum), we have already shown that Type Q periodicity is present.  These eight cases are not further distinguished by differing spectral periodicities; that is, we find no subtypes of Type Q.  To see this, we consider the fourfold-degenerate two-spinon multiplet with fractional crystal momenta $\boldsymbol{\kappa}$, $\boldsymbol{\kappa}'$.  First, we note that shifting $\kappa_x$ or $\kappa_y$ by $\pi$ merely permutes the four crystal momenta appearing in the multiplet.  Next, we can transform to another degenerate multiplet by acting on one spinon with $P^s_x$, $P^s_{xy}$, or ${\cal T}^s$.  These generators transform $\boldsymbol{\kappa}$ just as they do $\bq$, but with some additional shifts of $\kappa_x$, $\kappa_y$ by $\pi$, depending the $\sigma_i$'s.  Since these shifts do not affect the momenta of the two-spinon multiplet, they have no effect on the spectral periodicity.

Moving on to the eight cases $\sigma_{txty} = 1$, we consider
\beq
S^s T^s_\mu S^{s-1} = \sigma_{tS}
T_\mu^{s-1}, \quad  S \equiv \mathcal{T} (P_x P_{xy})^2 ,
\eeq
where $\sigma_{tS} = \sigma_{typx} \sigma_{txpx} \sigma_{Ttx}$.  The symmetry $S$ combines spatial inversion with time reversal to leave $\bfm{q}$ invariant.  Applying $S^s(2)$ to the state $\ket{a} = \ket{\bq_a ; z_a}$ gives
\beq
T_\mu \left[ S^s(2) \ket{a} \right] 
= \sigma_{tS} e^{iq_\mu} \left[ S^s(2) \ket{a} \right] ,
\eeq
so that when $\sigma_{tS} = -1$ there are degenerate states at $\bfm{q}$ and $\bfm{q} + (\pi,\pi)$.  We have thus shown that spectral periodicity of Type D is present in the four cases where $\sigma_{txty} = 1$ and $\sigma_{tS} = -1$.  Using arguments similar to that given above, it can be shown that there are no subtypes of Type D~\cite{note1}.  

The four cases with $\sigma_{txty} = \sigma_{tS} = 1$ remain.  We have
\beq
T_y \left[ P_x^s(2) \ket{a} \right] 
= \sigma_{typx} e^{iq_y} \left[ P_x^s(2) \ket{a} \right] ,
\eeq
so that, when $\sigma_{typx} = -1$, the degenerate states $\ket{a}$ and $P_x^s(2) \ket{a}$ have $y$-component of crystal momentum $q_y$ and $q_y + \pi$, respectively.  However, there is no simple relationship between the $x$-components of crystal momentum for these two states.  The degeneracy of these two states is thus not transparent in ${\cal N}(\bq, \omega)$, but is exposed in the reduced density of states ${\cal N}_y(q_y,\omega)$, and we have Type D$1d$ spectral periodicity.  

\begin{figure}
\subfigure[]{ \label{fig:doubled}
\includegraphics[width=1in]{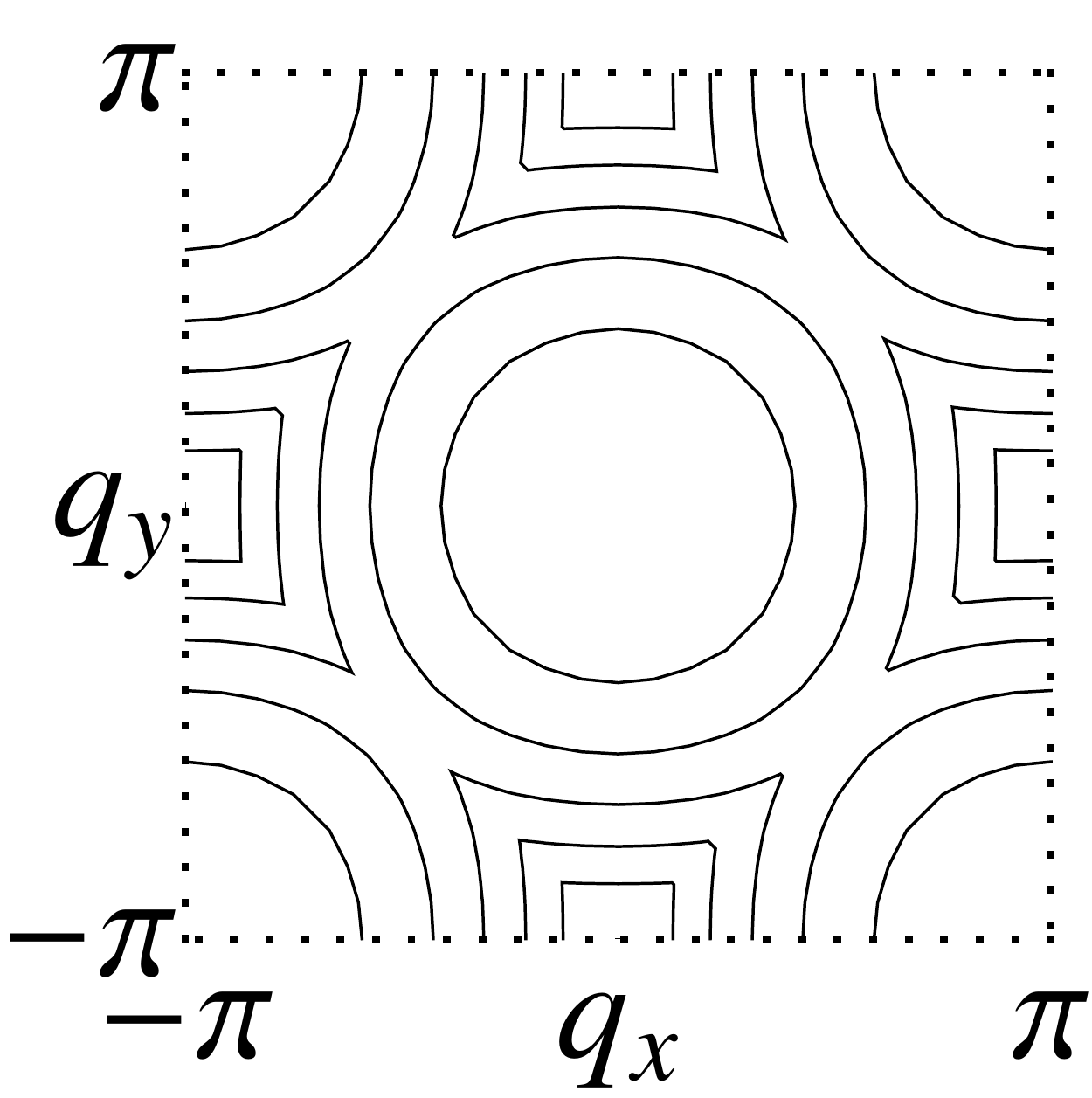} 
}
\subfigure[]{ \label{fig:leastsym}
\includegraphics[width=1in]{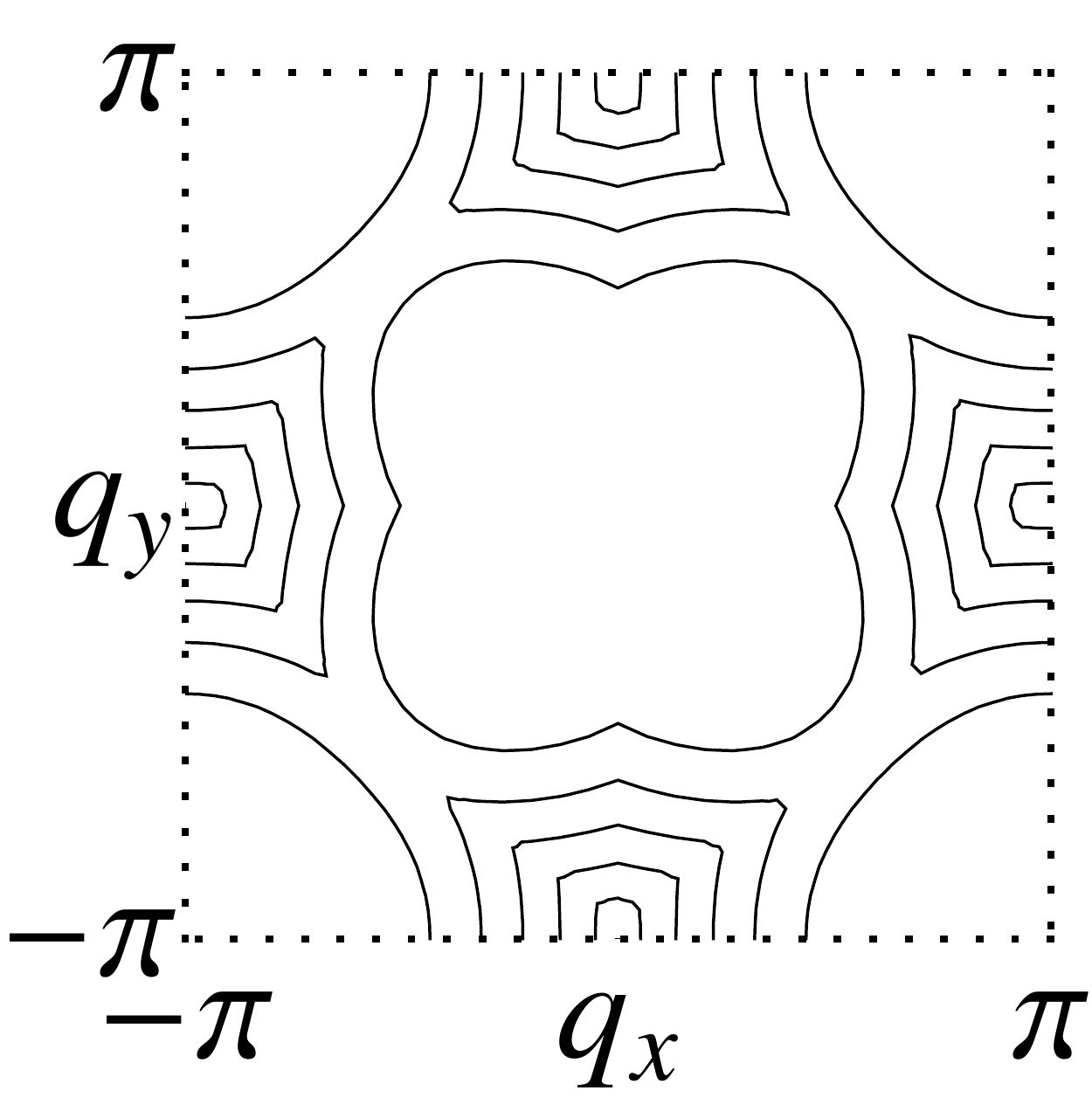}
}
\subfigure[]{ \label{fig:leastsymproj}
\includegraphics[width=1in]{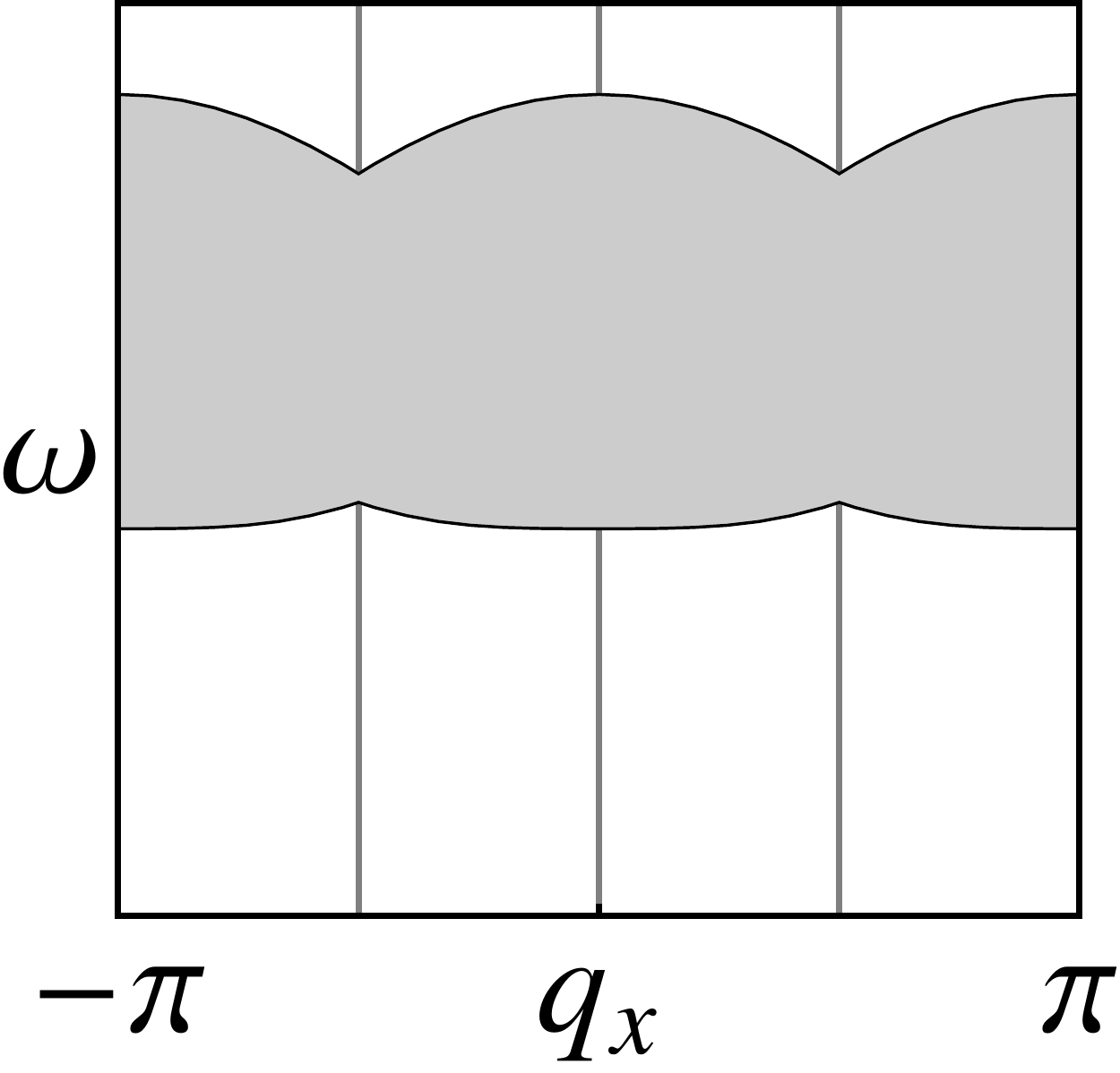}
}
\caption{(a) Contour plot of a representative mean-field $\Omega(\bfm{q})$ for State-D, with  $\Omega(\bfm{q}) = \Omega[\bfm{q} + (\pi,\pi)]$.  (b) Representative mean-field $\Omega(\bfm{q})$ for State-D$1d$, which shows no apparent enhanced  spectral periodicity.  (c) Plot of ${\cal N}_x(q_x, \omega)$ for State-D$1d$; the shaded region indicates $(q_x, \omega)$ for which ${\cal N}_x(q_x, \omega) = {\cal N}_x(q_x + \pi, \omega)$ is non-zero.}
\end{figure}

The two cases with  D$1d$ spectral periodicity satisfy $\sigma_{txty} = 1$, and $\sigma_{typx} = \sigma_{txpx} \sigma_{T tx} = -1$.  There are no sub-types of Type D$1d$, because these two cases can be mapped onto one another.  To see this, we consider the action of symmetries on $\bfm{k}$:
\begin{alignat}{2}
\mathcal{T}^s: & & \quad& (k_x,k_y) \rightarrow 
(-k_x,-k_y) + s_{Ttx} (\pi,\pi) \notag\\
P_{xy}^s: & & \quad& (k_x,k_y) \rightarrow (k_y,k_x) \notag\\
P_x^s: & & & (k_x,k_y) \rightarrow 
(-k_x,k_y) +  (s_{txpx} \pi, s_{typx} \pi) , \label{eqn:symmact}
\end{alignat}
where each $s_i = 0,1$, according to $\sigma_i = (-1)^{s_i}$.  
From this form, it is apparent that shifting the origin of $\bfm{k}$ by $(\pi/2,\pi/2)$ flips both $\sigma_{Ttx}$ and $\sigma_{txpx}$, leaving $\sigma_{typx}$ unchanged.  For two-spinon states, this shifts the origin of $\bq$ by $(\pi,\pi)$, but such a shift is innocuous since $\bq = (0,0)$ and $\bq = (\pi,\pi)$ are identical from a symmetry perspective.  In particular, flipping both $\sigma_{txpx}$ and $\sigma_{Ttx}$ does not affect the spectral periodicity.

The remaining two cases have $\sigma_{txty} = \sigma_{typx} = \sigma_{txpx} \sigma_{Ttx} = 1$. By shifting the origin as above, the case $\sigma_{Ttx} = \sigma_{txpx} = -1$ can be mapped onto the trivial case with all $\sigma_i = 1$.  In these cases we thus have a Type O spectrum, with no enhanced spectral periodicity.

To find Type D and D$1d$ states
we turn to the fermionic parton theory of spin liquids in $S = 1/2$ Heisenberg models \cite{wen02,gchen12}.  In this approach, one represents the spin at lattice site $\br$ in terms of a $S = 1/2$ fermion $f_{\alpha}(\bfm{r})$ ($\alpha = \uparrow, \downarrow$), with local constraint $\sum_{\alpha} f^\dagger_{\alpha}(\bfm{r}) f^{\vphantom\dagger}_{\alpha}(\bfm{r}) = 1$.  In mean-field theory the action of symmetry on fermions is described by a PSG~\cite{wen02}.  Depending on the structure of the mean-field saddle point, one obtains a gapped $\mathbb{Z}_2$ spin liquid upon incorporating fluctuations.  For such states, knowing the PSG, it is straightforward to compute the spinon fractionalization class.  Moreover, ${\cal N}(\bq, \omega)$ is easily computed within mean-field theory.

Within this framework, we construct two different gapped $\mathbb{Z}_2$ spin liquids with Type D and D$1d$ spectral periodicity. State-D has $\sigma_{txty} = 1$ and $\sigma_{txpx} = \sigma_{Ttx} = \sigma_{typx} = -1$, while State-D$1d$ has $\sigma_{txty} = \sigma_{Ttx} = 1$ and $\sigma_{txpx} = \sigma_{typx} = -1$.  Mean-field Hamiltonians and PSGs for each state are given in the Supplemental Material~\cite{note1}.  The mean-field $\Omega(\bq)$ is plotted for both states in Figs.~\ref{fig:doubled}, \ref{fig:leastsym}, while ${\cal N}_x(q_x,\omega)$ is shown in Fig.~\ref{fig:leastsymproj} for State-D$1d$, where the expected periodicities are evident.  

To summarize, we have shown that fractionalization of space group symmetry is manifest in the two-spinon spectrum, as a dramatic enhanced periodicity in the Brillouin zone. This periodicity is a direct, sharp consequence of the type of space group fractionalization, and is accessible to spectroscopic probes.  We therefore believe that this effect may play an important role in the experimental search for new quantum spin liquids, and perhaps more broadly.

\emph{Acknowledgements.}---We thank T.~Senthil for inspiring discussions and Dmitry Reznik for helpful comments. MH is grateful to Hao Song for a related collaboration. This work was supported by AFOSR/DARPA grant FA8750-12-2-0308 (AME) and by the David and Lucile Packard Foundation (MH and AME). The Institute for Quantum Information and Matter (IQIM) is an NSF Physics Frontiers Center with support from the Gordon and Betty Moore Foundation.

\bibliography{fracnumlett}


\newpage

\numberwithin{equation}{subsection}

\section*{Spectroscopic signatures of crystal momentum fractionalization: Supplemental Material}

{\center Andrew M.~Essin and Michael Hermele

\vspace{-0.14cm}

\subsection{Relations for square lattice space group and time reversal symmetry}
}

Following Ref.~\cite{Essin2013} of the main text [\href{http://dx.doi.org/10.1103/PhysRevB.87.104406 }{Phys.~Rev.~B 87, 104406 (2013)}, \href{http://arxiv.org/abs/1212.0593}{arXiv:1212.0593}], 
we list the 10 relations that define the symmetry group consisting of  square lattice space group and time reversal operations, in terms of the generators $T_x$, $P_x$, $P_{xy}$ and ${\cal T}$ defined in the main text (recall that $T_y = P_{xy} T_x P_{xy}$.)  The relations are:
\begin{eqnarray}
T_x T_y T^{-1}_x T^{-1}_y &=& 1 \\
T_x P_x T_x P^{-1}_x &=& 1 \\
T_y P_x T^{-1}_y P^{-1}_x &=& 1 \\
{\cal T} T_x {\cal T}^{-1} T_x^{-1} &=& 1 \\
P_x^2 = P_{xy}^2 = {\cal T}^2 &=& 1 \\
(P_x P_{xy})^4 &=& 1 \\
{\cal T} P_x {\cal T}^{-1} P^{-1}_x &=& 1 \\
{\cal T} P_{xy} {\cal T}^{-1} P^{-1}_{xy} &=& 1 \text{.}
\end{eqnarray}

\subsection{No subtypes of Type D spectral periodicity}

Here, we consider the four choices of $\sigma_i$ satisfying $\sigma_{txty} = 1$, $\sigma_{tS} = -1$, with Type D spectral periodicity.  We show that each of these four cases has the same spectral periodicity, and thus there are no subtypes of Type D.

Since $\sigma_{tS} = -1$, single-spinon eigenstates can be grouped into degenerate doublets with crystal momenta $\bfm{k}$ and $\bfm{k} + (\pi,\pi)$.  Therefore, each two-spinon state with single-spinon crystal momenta $\bfm{k}$, $\bfm{k}'$ is part of a degenerate quadruplet, in which the crystal momenta $\bq$ and $\bq + (\pi,\pi)$ each appear twice.  Making a shift $\bfm{k} \to \bfm{k} + (\pi,\pi)$ has no effect on the quadruplet.  

The four choices of the $\sigma_i$ under consideration can be specified by the ordered triple $(\sigma_{Ttx}, \sigma_{txpx},\sigma_{typx})$, and are
\begin{eqnarray}
{\rm D}_A &=& (-1,-1,-1) \notag\\
{\rm D}_B &=& (1,1,-1) \notag\\
{\rm D}_C &=& (-1,1,1) \notag\\
{\rm D}_D &=& (1,-1,1) \text{.}
\end{eqnarray}
First, we note that ${\rm D}_A$ and ${\rm D}_B$ can be mapped into one another by shifting the origin of $\bfm{k}$ by $(\pi/2,\pi/2)$, as discussed in the main text, and thus have the same spectral periodicity.  The same holds for ${\rm D}_C$ and ${\rm D}_D$.  It is thus sufficient to focus on ${\rm D}_A$ and ${\rm D}_C$.

We now consider the effect of acting with $P^s_x$, $P^s_{xy}$ and ${\cal T}^s$ on one spinon in the quadruplet described above.  This action is given in Eq.~(17) 
of the main text, and we see that for both ${\rm D}_A$ and ${\rm D}_C$, the only effect of the non-trivial $\sigma_i$'s is to augment the ordinary symmetry transformation of $\bfm{k}$ with a shift of $(\pi,\pi)$ in some cases.  For ${\rm D}_A$ this shift is present for both $P^s_x$ and ${\cal T}^s$, while for ${\rm D}_C$ it is only present for ${\cal T}^s$.  Because such shifts have no effect on the quadruplet, and because the presence/absence of the $(\pi,\pi)$ shift for $P^s_x$ is the only difference between ${\rm D}_A$ and ${\rm D}_C$, the spectral periodicity is the same in both cases.

\subsection{Parton mean-field $\mathbb{Z}_2$ spin liquid states}

Here, we use fermionic parton theory to exhibit mean-field, gapped $\mathbb{Z}_2$ spin liquid states with Type D and D$1d$ spectral periodicity.  We dub these State-D and State-D$1d$, respectively.  The PSG is specified by the action of symmetry generators on the two-component fermionic spinon field $\psi =  (f_{\up}, f_{\down}^\dag)^T$~[X.-G.~Wen, \href{http://dx.doi.org/10.1103/PhysRevB.65.165113}{Phys.~Rev.~B 65, 165113 (2002)}, \href{http://arxiv.org/abs/cond-mat/0107071}{arXiv:cond-mat/0107071}],
which can be chosen to be
\begin{align}
T_x^s \,\psi(x,y) &= \sigma_{txty}^{y} \psi(x+1,y) \notag\\
\mathcal{T}^s \,\psi(x,y) &= \sigma_{Ttx}^{x + y} (i\tau^y) g_T \psi(x,y) \notag\\
P_x^s \,\psi(x,y) &= \sigma_{txpx}^{x} \sigma_{typx}^{y} g_{P_x} \psi(-x,y) \notag\\
P_{xy}^s \,\psi(x,y) &= \sigma_{txty}^{x y} g_{P_{xy}} \psi(y,x).
\end{align}
Here $g_T$, $g_{P_x}$, and $g_{P_{xy}}$ are $2\times2$ matrices and $\tau^{x,y,z}$ are the $2 \times 2$ Pauli matrices.  It is straightforward to construct Hamiltonians quadratic in $\psi$ that are invariant under these transformations.

To construct State-D, we set $\sigma_{txty} = 1$, $\sigma_{txpx} = \sigma_{Ttx} = \sigma_{typx} = -1$, and construct a Bogoliubov--de Gennes Hamiltonian $\psi^\dag \mathcal{H}_D \psi$ that involves hopping and pairing terms to next-nearest-neighbor, with 
\begin{multline}
\mathcal{H}_D(\bfm{k}) = 
\left( u_0 + u_2 \cos k_x \cos k_y \right) \tau^x \\ + 
u_1 (\cos k_x + \cos k_y) \tau^z + u_2' \sin k_x \sin k_y \tau^y \text{.}
\end{multline}
In this PSG, $g_{P_{xy}} = \tau^0$ (the $2\times2$ identity matrix), $g_{P_x} = i\tau^x$, and $g_T = i\tau^z$.  The physical (two-spinon) spectrum is shown in Fig.~2(a)
of the main text for parameters $u_1 = u_0/4$, $u_2 = -u_0/2$, and $u_2' = u_0/4$.  As expected, it has Type D spectral periodicity, with equivalent minima at $(0,0)$ and $(\pi,\pi)$.  As discussed in the main text, when we include fluctuations about this mean-field state we arrive at a $\mathbb{Z}_2$ spin liquid which we term State-D.

To construct State-D$1d$, we modify the above PSG by setting $\sigma_{Ttx} = 1$.  Only pairing terms appear now, and a generic Hamiltonian with the same range is
\begin{multline}
\mathcal{H}_{D1d}(\bfm{k}) = \left( u_0 + u_2 \cos k_x \cos k_y \right) \tau^x \\ + 
\left[ u_1 (\cos k_x + \cos k_y) + u_2' \sin k_x \sin k_y \right] \tau^y .
\end{multline}
An example spectrum is shown in Figs.~2(b)
and 2(c)
of the main text, with the same parameters as for State-D.  As expected, the global minima at $(0,0)$ and $(\pi,\pi)$ are degenerate but inequivalent, and the projection of the density of states to $q_x$ is periodic under $q_x \to q_x + \pi$.

\end{document}